\documentclass[11pt]{article}
\title{Aggregated Sure Independence Screening for Variable Selection with Interaction Structures}
\author{Tonglin Zhang\footnote{Department of Statistics, Purdue University, 150 North University Street, West Lafayette, IN 47907-2067, Email: tlzhang@purdue.edu}}
\usepackage{xcolor}
\usepackage{subfigure}
\usepackage{epsfig}
\usepackage{mathrsfs}
\usepackage{amsfonts}
\usepackage{amsmath}
\usepackage{wrapfig,lipsum,booktabs}
\usepackage{bm}
\usepackage{algorithm}
\usepackage{algpseudocode}
\usepackage{pifont}
\def\qed{\hfill$\diamondsuit$}

\newtheorem{thm}{Theorem}

\newtheorem{lem}{Lemma}

\begin{document}
\maketitle
\def\eqalign#1{\null\,\vcenter{\openup\jot\ialign
              {\strut\hfil$\displaystyle{##}$&$\displaystyle{{}##}$
               \hfil\crcr#1\crcr}}\,}


\setcounter{page}{1}
\begin{abstract}
A new method called the aggregated sure independence screening is proposed for the computational challenges in variable selection of interactions when the number of explanatory variables is much higher than the number of observations (i.e., $p\gg n$). In this problem, the two main challenges are the strong hierarchical restriction and the number of candidates for the main effects and interactions. If $n$ is a few hundred and $p$ is ten thousand, then the memory needed for the augmented matrix of the full model is more than $100{\rm GB}$ in size, beyond the memory capacity of a personal computer.  This issue can be solved by our proposed method but not by our competitors. Two advantages are that the proposed method can include important interactions even if the related main effects are weak or absent, and it can be combined with an arbitrary variable selection method for interactions. The research addresses the main concern for variable selection of interactions because it makes previous methods applicable to the case when $p$ is extremely large.  
\end{abstract}



\section{Introduction}
\label{sec:introduction}

Variable selection of interactions is more difficult than main effects because it requires the strong hierarchical (SH) restriction. For instance, if an ultrahigh-dimensional linear model (HDLM) is considered such that the model can be expressed as
\begin{equation}
\label{eq:definition of the full interaction effect model} 
{\bm y}=\beta_0{\bm x}_0+\sum_{j=1}^p \beta_j{\bm x}_j+\sum_{j=1}^{p-1}\sum_{k=j+1}^p \beta_{jk}{\bm x}_j\circ {\bm x}_k+{\bm\epsilon}, {\bm\epsilon}\sim {\cal N}({\bm 0},\sigma^2{\bf I})
\end{equation}
with $p\gg n$, where ${\bm y}=(y_1,\dots,y_n)^\top\in\mathbb{R}^n$ is the response vector,  ${\bm x}_j=(x_{1j},\dots,x_{nj})^\top\in\mathbb{R}^n$ with ${\bm x}_0={\bm 1}$ is the $j$th explanatory variable, and ${\bm x}_j\circ{\bm x}_k=(x_{1j}x_{1k},\cdots,x_{nj}x_{nk})^\top$ with ${\bm x}_0\circ{\bm x}_k={\bm x}_k$ for $j=0$ is the Hadamard product (i.e., elementwise product) of ${\bm x}_j$ and ${\bm x}_k$. It is assumed that the response and explanatory variables are standardized such that they satisfy ${\bf 1}^\top{\bm y}={\bf 1}^\top{\bm x}_j=0$ and $\|{\bm x}_j\|^2=\|{\bm y}\|^2=n$ for all $j=1,\dots,p$. We still need $\beta_0$ in~\eqref{eq:definition of the full interaction effect model} because of the presence of the interactions. A challenge is the SH restriction, which means that the final selected model needs to include ${\bm x}_j$ and ${\bm x}_k$ if it contains ${\bm x}_j\circ{\bm x}_k$. Another is the number of candidates of the main effects and interactions. If $p\gg n$, then the number of the candidates for the main effects and interactions is $p(p+1)/2$, leading to computational difficulties in penalized least squares (PLS) or penalized maximum likelihood (PML) methods for variable selection.  To overcome the difficulty, a common way is to apply a sure independent screening (SIS) approach~\cite{fanlv2008}. The goal is to reduce the number of candidate variables to a small number, such that a variable selection method can be applied. Therefore, the two main difficulties in selecting interactions from~\eqref{eq:definition of the full interaction effect model} when $p$ is large are the SH restriction and the number of candidates of the main effects and interactions. Both are addressed in the research. 

We present our method based on the HDLM defined by~\eqref{eq:definition of the full interaction effect model}. The idea of our method can be extended to other types of ultrahigh-dimensional statistics, including the ultrahigh-dimensional generalized linear models (HDGLMs) for the case when the response follows an exponential family distribution. The details are displayed in Section \ref{sec:possible extension}. 

For selection of interactions,  it is necessary to account for well-established restrictions, such as heredity~\cite{chipman1996,hamada1992} and marginality~\cite{mccullagh2002,nelder1977,nelder1994}, interpreted by the SH restriction (also called the strong heredity). The SH restriction requires $\hat\beta_{jk}\not=0\Rightarrow \hat\beta_j\hat\beta_k\not=0$, where $\hat\beta_j$ and $\hat\beta_{jk}$ are the estimates of $\beta_j$ and $\beta_{jk}$ provided by a variable selection method, respectively. Violations of the SH restriction can only occur in special situations, implying that it should be treated as the default~\cite{bien2013}. To ensure that the SH restriction is satisfied, the earliest method was the strong heredity interaction model (\textsf{SHIM})~\cite{choi2010} which expresses interaction parameters as $\beta_{jk}=\gamma_{jk}\beta_j\beta_k$ for all $j\not=k$. Because the \textsf{SHIM} might penalize an important interaction to be $0$ if one of the related main effects has been penalized to be $0$, a method called the variable selection using adaptive nonlinear interaction structures in high dimensions (\textsf{VANISH})~\cite{radchenko2010} was proposed. Early methods also included the hierarchical net (\textsf{hierNet})~\cite{bien2013} and the group regularized estimation under structural hierarchy (\textsf{GRESH})~\cite{she2018}. The \textsf{hierNet} enforces $\sum_{k=1}^p |\beta_{jk}|\le |\beta_j|$ for all $j\in\{1,\dots,p\}$. The \textsf{GRESH}, which can be treated as a modification of the \textsf{VANISH}, uses two distinct formulations to penalize the main effects and interactions. A concern is that the computation of these methods is time-consuming even when $p$ is a few hundred. For instance, in one of our numerical studies, the computation for the \textsf{GRESH} and the \textsf{SHIN} when $n=200$ and $p=500$ took a few days. The \textsf{hierNet} failed when it was applied. We then propose the aggregated correlation sure independence screening (AcorSIS) method. Our research showed that the AcorSIS method reduced the computation of the \textsf{GRESH} and the \textsf{SHIN} from a few weeks to a few seconds even when $p$ was a few thousand. The AcorSIS can be combined with an arbitrary variable selection method for interactions. It ensures that the SH restriction is satisfied. Combined with the AcorSIS, any variable selection methods for interactions can be applied to the case when $p$ is large. We successfully address the two main difficulties in variable selections of interactions. 

Due to the computational difficulty of the \textsf{GRESH}, the \textsf{SHIN}, and the \textsf{hierNet}, recent work focuses on a two-step procedure. The first is the screening step. The goal is to reduce the computational burden of the penalization step (i.e., the second step). It is achieved by excluding unimportant main effects and interactions based on the magnitudes of a marginal or a conditional measure. The second is the penalization step. The goal is to select the important main effects and interactions based on the result provided by the screening step. Examples include the high-dimensional quadratic regression (\textsf{HiQR})~\cite{wang2023} and the sparse reluctant interaction modeling (\textsf{sprinter})~\cite{yu2019}. In the two methods, we find a concern that screening is carried out by the all-pairs SIS~\cite{hallxue2014}, leading to violations of the SH restriction. The \textsf{RAMP}~\cite{hao2018} selects the main effects first and then captures the interactions in the second for the SH restriction. We find a concern that the \textsf{RAMP} is unlikely to capture an important interaction if the related main effects are weak or absent. Although the concern is partially addressed by the \textsf{hierScale}~\cite{hazimeh2020}, it is still hard to include an important interaction if the related main effects are weak or absent. This issue is addressed if the proposed AcorSIS is combined with the \textsf{GRESH}. 

The main contribution of our research is that we classify screening for interactions into two categories: variable-based screening and effect-based screening. The proposed AcorSIS is an example of variable-based screening. The previous all-pair SIS is an example of effect-based screening. In particular, a variable-based screening method searches for each ${\bm x}_j$ and claims ${\bm x}_j$ as an active variable if the main effect of ${\bm x}_j$ is important or there exists $k$ such that the $(j,k)$th interaction is important. It can address the three scenarios when ${\bm x}_j\circ{\bm x}_k$ is important in the true model. In the first, both main effects ${\bm x}_j$ and ${\bm x}_k$ are important. In the second, one of the main effects ${\bm x}_j$ and ${\bm x}_k$ is important but the other is not. In the third, none of the main effects ${\bm x}_j$ and ${\bm x}_k$ are important. A variable-based screening method likely chooses both ${\bm x}_j$ and ${\bm x}_k$ as candidates of variables in all three scenarios. An effect-based screening method likely keeps ${\bm x}_j\circ{\bm x}_k$ but not ${\bm x}_j$ or ${\bm x}_k$ in the third scenario, leading to a violation of the SH restriction. Thus, variable-based screening is more appropriate than effect-based screening. 

 The minor contribution of our research is that our result shows that the performance of the combination of the AcorSIS and the \textsf{GRESH}, called the modified \textsf{GRESH} in this research, is better than many recently developed methods. Examples include the \textsf{HiQR}, the \textsf{sprinter}, the \textsf{RAMP}, and the \textsf{hierScale}. The computational issue of the previous \textsf{GRESH} is overcome by the modified \textsf{GRESH}. We then recommend it for selections of interactions when $p$ is large. 

In the literature, screening was originally proposed for the main effects only. Examples include the iterative sure independence screening (SIS)~\cite{fanlv2008,fansamworth2009,fansong2010}, the conditional SIS~\cite{barutfan2016}, the tilting screening~\cite{halltitterington2009}, the generalized correlation screening~\cite{hallmiller2009}, the nonparametric screening~\cite{fanfeng2011}, the robust rank correlation-based screening~\cite{lipeng2012}, the model-free feature screening~\cite{cuili2015,zhuli2011}, the martingale difference correlation screening~\cite{shaozhang2014}, the fused Kolmogorov filter screening (Kfilter)~\cite{mai2015},  the pairwise SIS~\cite{panwang2016}, the PC-screening~\cite{liuke2022}, the weighted leverage score (WLS) screening~\cite{zhongliu2021}, the category adaptive screening (CAS)~\cite{xielin2020}, and the concordance index SIS (CSIS)~\cite{chengwang2023}. 

Recently, the importance of interaction screening was noticed. A few interaction screening methods were proposed. Examples include the all-pairs SIS~\cite{hallxue2014}, the distance correlation SIS (DCSIS)~\cite{lizhong2012}, the sliced inverse regression for interaction detection (SIRI)~\cite{jiangliu2014}, the interaction pursuit via distance correlation (IPDC)~\cite{kongli2017}, the ball correlation SIS (BcorSIS)~\cite{panwang2019}, the BOLT-SSI~\cite{zhoudai2022}, and the interaction forward method (\textsf{iForm})~\cite{haozhang2014}. We examine all of these and find a few concerns. For instance, the all-pairs SIS adopted by the \textsf{HiQR}~\cite{wang2023} and \textsf{sprinter}~\cite{yu2019} violates the SH restriction. The DCSIS and MDCSIS use the distance correlation and martingale difference correlation, respectively, but not the correlation. They are also developed from a marginal measure for the main effects. The IPDC uses the distance correlation between ${\bm y}^2$ and ${\bm x}_j^2$ to screen interactions, but they are still based on marginal measures for the main effects. A similar issue appears in the BcorSIS. The BOLT-SSI is developed based on the techniques of contingency tables, which needs a discretization of variables in its implementation. The \textsf{iForm} is developed under the strong heredity assumption of the true model. It is hard to capture an important interaction if none of the related main effects is important. Our AcorSIS does not suffer from these concerns, and it is better than the previous SIS methods for interactions. 

The article is organized as follows.  In Section~\ref{sec:method}, we propose our method. In Section~\ref{sec:theoretical property}, we derive the theoretical properties of our method under a set of suitable regularity conditions. In Section~\ref{sec:simulation}, we compare our method with our competitors by simulations. In Section~\ref{sec:application}, we apply our method to a real data example. We provide a discussion in Section \ref{sec:discussion}. We put all of the proofs in the appendix.

\section{Method}
\label{sec:method}

We introduce our method in Section~\ref{subsec:aggregated correlation learning for screening Interactions}. We evaluate the computational issues in Section~\ref{subsec:memory requirement and computational efficiency}. We display a few possible extensions in Section~\ref{sec:possible extension}.

\subsection{Aggregated Correlation Learning for Screening Interactions}
\label{subsec:aggregated correlation learning for screening Interactions}

The full model~\eqref{eq:definition of the full interaction effect model} has $p$ variables, $p$ main effects, and $p(p-1)/2$ interactions. They are candidates considered by a variable selection method. It is assumed that the true main effects and interactions are sparse. Only a few of $\beta_j^*$ and $\beta_{jk}^*$ are nonzero, where $\beta_j^*$ represents the true main effect of ${\bm x}_j$ and $\beta_{jk}^*$ represents the true interactions of ${\bm x}_j\circ{\bm x}_k$ in~\eqref{eq:definition of the full interaction effect model}. The true model is 
\begin{equation}
\label{eq:true model}
{\bm y}=\beta_0^*{\bm x}_0+\sum_{j\in{\cal T}_M}\beta_j^*{\bm x}_j+\mathop{\sum\sum}_{(j,k)\in{\cal T}_I}\beta_{jk}^* {\bm x}_j\circ{\bm x}_k+{\bm\epsilon}^*, {\bm\epsilon}^*\sim{\cal N}({\bm 0},\sigma^{*2}{\bf I}),
\end{equation}
where $\beta_j^*\not=0$ iff $j\in{\cal T}_M\subseteq\{1,\dots,p\}$, $\beta_{jk}^*\not=0$ iff $(j,k)\in{\cal T}_I\subseteq\{(j,k):1\le j<k\le p\}$, ${\cal T}_M$ represents the set of the important main effects, and ${\cal T}_I$ represents the set of the important interactions. The true (i.e. active) variable set is ${\cal T}=\{j: j\in{\cal T}_M~{\rm or}~\exists k\not=j~{\rm such~that}~(j,k)\in{\cal T}_I\}$. The true model has $|{\cal T}|$ variables, $|{\cal T}_M|$ main effects, and $|{\cal T}_I|$ interactions. Following the literature~\cite{hamada1992}, the SH restriction is required in the selected model even though it is not present in the true model. If the SH restriction is satisfied in~\eqref{eq:true model}, then ${\cal T}={\cal T}_M$; otherwise ${\cal T}\supset{\cal T}_M$. 

The related main effects of ${\bm x}_{j}\circ{\bm x}_k$ are ${\bm x}_j$ and ${\bm x}_k$. We say that ${\bm x}_j$ is an important main effect if $\beta_j^*\not=0$ and ${\bm x}_j\circ{\bm x}_k$ is an important interaction if $\beta_{jk}^*\not=0$. We say that ${\bm x}_j$ is an active variable if ${\bm x}_j$ is an important main effect or related to an important interaction. We say that ${\bm x}_j$ is a redundant variable if it is not an active variable. We study the case when a two-stage procedure is used under the framework of variable-based screening. The screening stage removes most of the redundant variables. The penalization stage selects the main effects and interactions based on the variable set provided by the screening stage. All the active variables should be kept in the screening stage; otherwise, at least one of the important main effects or interactions cannot be selected in the penalization stage. 

The goal of our research is to provide a shrunk variable set ${\cal S}\subset\{1,\dots,p\}$ with $|{\cal S}|\ll n$, such that ${\rm Pr}({\cal T}\subset{\cal S})$ is large. We restrict the penalization stage to variables in ${\cal S}$ only. Because $|\cal S|$ is small, the computational challenge in the penalization stage is solved. To achieve this goal, we focus on the derivation of a shrunk set of variables but not a shrunk set of effects. Therefore, we treat our method as a variable-based screening method. To achieve the research goal, we define the {\it aggregated correlation} (acor) coefficient for the $j$th variable as
\begin{equation}
\label{eq:definition of acor}
{\rm acor}({\bm x}_j)=\max_{k\in\{0,1,\dots,p\},k\not=j}|{\rm cor}({\bm x}_j\circ {\bm x}_k,{\bm y})|,
\end{equation} 
where ${\rm cor}({\bm x}_j\circ{\bm x}_0,{\bm y})={\rm cor}({\bm x}_j,{\bm y})$. Our numerical studies show that our method can capture all the active variables in general but our competitors cannot (See Section~\ref{sec:simulation} for details).

We expect that ${\rm acor}({\bm x}_j)$ is large if ${\bm x}_j$ is an active variable; or small otherwise. We provide a shrunk subset of variables as
\begin{equation}
\label{eq:rank of the acor}
\hat{\cal S}_\gamma=\{j: {\rm acor}({\bm x}_j)~{\rm  is~among~the~first}~d_\gamma~{\rm largest~of~all}\},
\end{equation}
where $d_\gamma=[{\gamma}n]$ denotes the integer part of ${\gamma}n$. The convenient value is $\gamma=1/\log{n}$, as it has been widely used previously~\cite[e.g.]{kongli2017}. We then implement a variable selection method for interactions with the shrunk model as
\begin{equation}
\label{eq:shrunk model}
{\bm y}=\beta_0{\bm x}_0+\sum_{j\in \hat{\cal S}_\gamma}\beta_j{\bm x}_j+\mathop{\sum\sum}_{j,k\in\hat{\cal S}_{\gamma},j<k}\beta_{jk}{\bm x}_j\circ{\bm x}_k+{\bm\epsilon},{\bm\epsilon}\sim {\cal N}({\bm 0},\sigma^2{\bf I}).
\end{equation}
Because ${\rm Pr}({\cal T}\subseteq\hat{\cal S}_\gamma)\approx1$, it is suitable to restrict the penalization stage on~\eqref{eq:shrunk model}. 

The AcorSIS reduces the number of candidate variables from $p$ in the full model to $d_\gamma$ in the shrunk model.  In practice, we may use a larger $\gamma$ to reduce the value of $d_\gamma$ if the convenient value of $\gamma$ is not adopted. Because the goal of the screening stage is not to miss any active variables, it is not necessary to make $d_{\gamma}$ very low. 

The penalization stage selects the main effects and interactions from~\eqref{eq:shrunk model}. Based on the shrunk model, the goal is to keep all the important and remove all the unimportant main effects and interactions. The shrunk model given by~\eqref{eq:shrunk model} has $d_\gamma$ candidate variables with $d_\gamma$ candidates for the main effects and $d_\gamma(d_\gamma-1)/2$ candidates for the interactions. For instance, if $\gamma=\log{n}$, $n=200$, and $p=10^4$, then $d_\gamma =[n/\log{n}]=37$. The number of candidates for the main effects and interactions is $740$, implying that a penalization method can be easily implemented. We then propose Algorithm~\ref{alg:aggregated correlation coefficient for interaction effects}.

\begin{algorithm}
\caption{\label{alg:aggregated correlation coefficient for interaction effects} Two-stage selection for the main effects and interactions from~\eqref{eq:definition of the full interaction effect model} using the AcorSIS }
\begin{algorithmic}[1]
\Statex{{\bf Input}: Response ${\bm y}\in\mathbb{R}^y$ and ${\bf X}=[{\bm x}_1,\dots,{\bm x}_p]\in\mathbb{R}^{n\times p}$}
\Statex{{\bf Output}: Important main effects and interactions and their estimates}
\Statex{\it Screening Stage}
\State{Compute ${\rm acor}({\bm x}_j)$ for every $j\in\{1,\dots,p\}$ by~\eqref{eq:definition of acor}}
\State{Calculate the shrunk subset of variables $\hat{\cal S}_\gamma$ by~\eqref{eq:rank of the acor} based on some $\gamma\in(0,1)$}
\Statex{\it Penalization Stage}
\State{Construct the shrunk model~\eqref{eq:shrunk model} by $\hat{\cal S}_\gamma$}
\State{Select the main effects and the interactions from~\eqref{eq:shrunk model}}
\Statex{\it End Iteration}
\State {Output}
\end{algorithmic}
\end{algorithm}

Algorithm~\ref{alg:aggregated correlation coefficient for interaction effects} has two stages. The screening stage removes most of the redundant variables. Because only variables contained by $\hat{\cal S}_{\gamma}$ are considered, the penalization stage cannot recover any important main effects or interactions if corresponding active variables are removed in the screening stage. Therefore, we need to ensure that ${\rm Pr}({\cal T}\subset \hat {\cal S}_\gamma)$ is high. It is not critical if $\hat{\cal S}_\gamma$ contains a few redundant variables. Theoretically, any PLS method can be treated as a candidate for the penalization stage.  In this research, we study the combinations of our AcorSIS method with the previous \textsf{GRESH}~\cite{she2018}, \textsf{SHIM}~\cite{choi2010}, and \textsf{hierNet}~\cite{bien2013} methods, respectively. The previous versions of the three methods have difficulties if $p$ is a few hundred. Our numerical studies show that the computation may take a few days even when $p$ is a few thousand. 

All of the \textsf{GRESH}, the \textsf{SHIM}, and the \textsf{hierNet} can be treated as extensions of the LASSO~\cite{tibshirani1996}. The \textsf{GRESH} and the \textsf{hierNet} can also be treated as modifications of the \textsf{VANISH}~\cite{radchenko2010}. An obvious concern of the \textsf{hierNet}~\cite{bien2013} is that it enforces a magnitude constraint as $\sum_{k\not=j}|\beta_{jk}| \le |\beta_j|$ to make the SH restriction satisfied. The \textsf{GRESH} does not have such a concern. The authors pointed out that the \textsf{GRESH} outperforms the \textsf{hierNet}. The conclusion was based on a simulation with $p=100$. To make the \textsf{GRESH} applicable when $p$ is a few thousand or even higher, we modify the objective function of \textsf{GRESH} for a set of variables ${\mathcal A}\subseteq\{1,\dots,p\}$ as
\begin{equation}
\label{eq:GRESH Lq}
\eqalign{
{\mathcal L}_{\lambda_1,\lambda_2,r}^{GRESH}({\bm\beta})&={1\over 2}\|{\bm y}-\beta_0{\bm x}_0-\sum_{j\in{\mathcal A}}{\bm x}_j\beta_j-\mathop{\sum\!\sum}_{j,k\in{\mathcal A},j<k}{\bm x}_j\circ{\bm x}_k\beta_{jk} \|^2 \cr
&+n\lambda_2\sum_{j\in{\mathcal A}} ( |\beta_j|^r+\sum_{k=1,k\not=j}^p |\beta_{jk}|^r)^{1/r}+n\lambda_1\mathop{\sum\!\sum}_{j,k\in{\mathcal A},j<k} |\beta_{jk}|.\cr
}
\end{equation}
The objective function given by~\eqref{eq:GRESH Lq} becomes the version proposed by the authors if we choose ${\mathcal A}=\{1,\dots,p\}$. The modified \textsf{GRESH} is derived if we choose ${\mathcal A}=\hat{\mathcal S}_\gamma$ in~\eqref{eq:GRESH Lq}. To implement the modified \textsf{GRESH}, we adopt the option recommended by the authors as $\lambda_2=0.5\lambda_1$ and $r=2$. Optimization of the \textsf{GRESH} proposed by the authors is developed by the alternating direction method of multipliers (ADMM) algorithm. A concern is that the ADMM has an additional parameter not contained in the objective function. To address this issue, we propose a coordinate descent algorithm. The details are displayed in Appendix~\ref{sec:coordinate descent or the gresh}. 

The \textsf{SHIM} is developed under the LASSO with an expression of $\beta_{jk}$ as $\gamma_{jk}\beta_j\beta_k$ for the SH restriction with the objective function as
\begin{equation}
\label{eq:objective function of SHIM}
\eqalign{
{\mathcal L}_{\lambda_1,\lambda_2}^{SHIM}({\bm\beta})=&{1\over 2}\|{\bm y}-\beta_0{\bm x}_0-\sum_{j\in{\mathcal A}}{\bm x}_j\beta_j-\mathop{\sum\!\sum}_{j,k\in{\mathcal A},j<k}{\bm x}_j\circ{\bm x}_k\gamma_{jk}\beta_j\beta_k\|^2\cr
&+n\lambda_2\sum_{j\in{\mathcal A}} |\beta_j|+n\lambda_1\mathop{\sum\!\sum}_{j,k\in{\mathcal A},j<k} |\gamma_{jk}|.
}
\end{equation}
The authors recommend $\lambda_1=\lambda_2$ in~\eqref{eq:objective function of SHIM}. Optimization is carried out by coordinate descent.

We find concerns of the \textsf{SHIM} and the \textsf{GRESH}. When the main effect ${\bm x}_j$ is weak or absent but the interaction ${\bm x}_j\circ{\bm x}_k$ is strong, the \textsf{SHIM} is likely to penalize $\beta_j$ to be $0$, leading to $\beta_{jk}=0$. This is not an issue in the \textsf{GRESH}. The \textsf{SHIM} satisfies: $\hat\beta_j=0~{\rm or}~\hat\beta_k=0\Rightarrow\hat\beta_{jk}=0$. The \textsf{GRESH} satisfies: $\hat\beta_{jk}\not=0\Rightarrow \hat\beta_j,\hat\beta_k\not=0$. The two properties are not equivalent in computations. 

We compare our proposed AcorSIS method with the built-in screening methods used in the previous \textsf{HiQR}~\cite{wang2023}, \textsf{sprinter}~\cite{yu2019}, \textsf{RAMP}~\cite{hao2018}, \textsf{SODA}~\cite{li2019}, and \textsf{hierScale}~\cite{hazimeh2020}. Our Monte Carlo simulations show that our method outperforms these methods in general. The details are displayed in Section~\ref{sec:simulation}.

Although the AcorSIS method is developed for the HDLM defined by~\eqref{eq:definition of the full interaction effect model}, it can also be implemented in HDGLMs when the response follows an exponential family distribution. The idea is to use the approach by~\cite{fanlv2008} for a logistic linear model with $n_1$ samples from class $1$ and $n_2$ samples from class $0$, where the response satisfies $y_i=1$ or $y_i=0$ for all $i=1,\dots,n$. We define the correlation coefficient as 
\begin{equation}
\label{eq:definition of correlation for 0 or 1}
\eqalign{
&{\rm cor}({\bm x}_j\circ{\bm x}_k,{\bm y})\cr
=&{{n-n_1\over n}\sum_{\{i: y_i=1\}} (x_{ij}x_{ik}-\overline {x_{j}x_{k}})-{n_1\over n}\sum_{\{i: y_i=0\}} (x_{ij}x_{ik}-\overline {x_{j}x_{k}})\over [\sum_{i=1}^n (x_{ij}x_{ik}-\overline{x_{j}x_{k}})^2( n_1-2n_1/n+ n_1^2/n)]^{1/2}},
}\end{equation}
 where $\overline {x_{j}x_{k}}$ is the sample mean of ${\bm x}_j\circ{\bm x}_k$. Therefore, our method is not limited to the HLDMs for normal responses. 

We treat our method as variable-based aggregated correlation learning because it relies on the aggregated correlation coefficient defined by~\eqref{eq:definition of acor}. Our method provides a shrunk set of variables but not a shrunk set of effects. The ${\rm acor}({\bm x}_j)$ value reflects the importance of variables. This is different from the previous correlation learning. If only main effects are studied, then we can assign ${\rm cor}({\bm x}_j\circ {\bm x}_k,{\bm y})=0$ for every $k\not=0$, $k\not=j$, leading to ${\rm acor}({\bm x}_j)={\rm cor}({\bm x}_j,{\bm y})$. Thus, the previous correlation learning proposed by~\cite{fanlv2008} is a special case of our method.

We compare our AcorSIS with the previous all-pair SIS proposed by \cite{hallxue2014}. The difference is that the all-pair SIS ranks the absolute values of ${\rm cor}({\bm x}_j\circ{\bm x}_k,{\bm y})$ for all distinct $j,k\in\{0,1,\dots,p\}$, leading to a shrunk sunset of effects as
\begin{equation}
\label{eq:shrunk set of effects all pair SIS}
\hat{\cal E}_\gamma=\{(j,k): |{\rm cor}({\bm x}_j\circ{\bm x}_k,{\bm y})|~{\rm is~among~the~first}~d_\gamma~{\rm largest~of~all}\},
\end{equation}
where $d_\gamma$ is the same as that defined in~\eqref{eq:rank of the acor}. After $\hat{\cal E}_\gamma$ is derived, the penalization stage selects the main effects and interactions from the shrunk model as
\begin{equation}
\label{eq:shrunk model all pair SIS}
{\bm y}={\bm\beta}_0{\bm x}_0+\sum_{(0,j)\in\hat{\cal E}_\gamma} \beta_j{\bm x}_j+\sum_{(j,k)\in\hat{\cal E}_\gamma,0<j<k}\beta_{jk}{\bm x}_j\circ{\bm x}_k+{\bm\epsilon},{\bm\epsilon}~{\cal N}(0,\sigma^2{\bf I}).
\end{equation}
A concern is that the all-pair SIS may miss ${\bm x}_j$ or ${\bm x}_j$ even if ${\bm x}_j\circ{\bm x}_k$ is caught in $\hat{\cal E}_\gamma$, leading to a violation of the SH-restriction. We discover this in our simulations.  

\subsection{Memory Requirement and Computational Efficiency}
\label{subsec:memory requirement and computational efficiency}

 It is important to evaluate the computational properties of a statistical method to ensure that it can be applied. The two commonly used measures are memory requirement in size and computational efficiency in floating operations. If the size of the memory requirement exceeds the size of the memory capacity of the computing system, then the implementation will be terminated. If the number of floating operations is large but the size of the memory requirement is under the limit, then the method can still be implemented. Thus, it is more important to evaluate the size of the memory requirement of a statistical method. 

The memory size of a floating number is $8$ bytes. To load ${\bm y}$ and ${\bf X}=[{\bm x}_1,\dots,{\bm x}_p]$, a computer needs to allocate $8n(p+1)$ bytes in memory. Because ${\rm acor}({\bm x}_j)$ can be calculated based on individual ${\rm cor}({\bm x}_j\circ{\bm x}_k,{\bm y})$, the derivation of ${\rm acor}({\bm x}_j)$ only needs to allocate $8n$ bytes in memory. The result can be stored in a vector with $p$ components, which needs $8p$ bytes in memory. The ranking of ${\rm acor}({\bm x}_j)$ for $j=1,\dots,p$ needs $8p$ bytes in memory. The minimum memory requirement for the screening stage of our method displayed in Steps 1 and 2 of Algorithm~\ref{alg:aggregated correlation coefficient for interaction effects} is $8n(p+1)+8n+8p+8p\le 8n(p+3)$ bytes in size. It is almost identical to the memory needed in size for loading the data. Therefore, the memory requirement of our method is low. After the screening stage is over, the penalization stage needs to allocate a matrix with $n$ rows and $(d_{\gamma}^2+d_{\gamma})/2+1$ columns. Penalization for variables in $\hat{\cal S}_\gamma$ needs to allocate at least three more matrices with the same size. The memory requirement is about $32n[(d_{\gamma}^2+d_{\gamma})/2+1]$ bytes, which is affordable. For instance, if $n=200$ and $p=10^4$, then the memory requirement of the screening stage of our method is about $16{\rm MB}$. If $d_{\gamma}=[n/\log(n)]=37$ is used, then the memory requirement of the penalization stage of our method is about $4.3{\rm MB}$. To compare, if the previous ranking all interactions method proposed by~\cite{hallxue2014} and recently used by~\cite{li2022,yu2019} is adopted, then the computation needs to allocate at least $3{\rm GB}$ memory in size. If a penalization method is directly applied to raw data, then the computer needs to allocate four matrices with $n$ rows and $(p^2+p)/2+1$ columns, which is about $300{\rm GB}$ memory in size. As the memory requirement in size of many well-established \textsf{R} and \textsf{python} packages is not optimized, the computer usually needs more memory capacity to implement these methods. 

For computational efficiency, the derivation of each ${\rm cor}({\bm x}_j\circ{\bm x}_k,{\bm y})$ needs $4n$ floating operations. The derivation of each ${\rm acor}({\bm x}_j)$ needs $4np$ floating operations. The derivation of all ${\rm acor}({\bm x}_j)$ for $j=1,\dots,p$ needs $O(4np^2)$ floating operations. If $n=200$ and $p=10^4$, then the implementation of the screening stage needs $8(10^{10})$ floating operations. This would take less than $10$ seconds if the computation were based on \textsf{C++} or \textsf{JAVA}. If the computation were based on \textsf{R}, then it would take a few minutes as \textsf{R} can be $500$ times slower than \textsf{C++}~\cite{aruoba2015}. It is still acceptable for real-world data applications.

\subsection{Possible Extension}
\label{sec:possible extension}

Although our method is developed based on the aggregation correlation coefficient, the idea can be extended to other aggregated coefficients. For instance, we can define the aggregated distance correlation coefficient by ${\rm adcor}({\bm x}_j,{\bm y})=\max_{k\in\{1,\dots,p\},k\not=j}|{\rm dcor}({\bm x}_j\circ{\bm y}_k,{\bm y})|$, where ${\rm dcor}({\bm u},{\bm v})$ is the distance correlation coefficient between vectors ${\bm u}$ and ${\bm v}$. The distance correlation coefficient has been previously used~\cite{lizhong2012,kongli2017}, but the aggregated distance correlation has not. The computation of the distance correlation coefficient between two random vectors has been incorporated in the \textsf{dcorT} function of the package \textsf{energy} of \textsf{R}. We can also define the aggregated generalized correlation coefficient  by ${\rm agcor}({\bm x}_j,{\bm y})=\max_{k\in\{1,\dots,p\},k\not=j}|{\rm gcor}({\bm x}_j\circ{\bm y}_k,{\bm y})|$, where ${\rm gcor}({\bm u}, {\bm v})={\rm cor}[h({\bm u}),{\bm v}]$ is the generalized correlation coefficient between ${\bm u}$ and ${\bm v}$ based on transformation $h(\cdot)$ on ${\bm u}$ used previously by~\cite{hallxue2014}. Our idea can be extended to HDGLMs if the likelihood ratio statistic is used. The approach is to define an aggregated likelihood ratio statistic as 
\begin{equation}
\label{eq:aggregated likelihood ratio}
{\rm a}\Lambda({\bm x}_j)=\max_{k\in\{0,1,\dots,p\},k\not=j} \Lambda({\bm x}_j\circ{\bm x}_k),
\end{equation}
where $\Lambda({\bm x}_j\circ{\bm x}_k)$ is the likelihood ratio statistic for the difference between the null GLM (i.e., the model with intercept only) and the GLM with ${\bm x}_j\circ{\bm x}_k$ as the only explanatory variable. Note that ${\rm a}\Lambda({\bm x}_j)$ defined by~\eqref{eq:aggregated likelihood ratio} is not limited to HDGLMs. The basic finding of this research can be extended to a wide range of statistical models for the screening stage of variable selection of interactions.

\section{Theoretical Property}
\label{sec:theoretical property}

We evaluate the asymptotic properties of our method when $n\rightarrow\infty$ with $p\rightarrow\infty$ simultaneously, where $p$ can grow exponentially fast with $n$. The main issue is to show $\lim_{n\rightarrow\infty} {\rm Pr}({\cal T}\subseteq\hat{\cal S}_\gamma)=1$ under a few suitable regularity conditions. We can modify the standard method utilized by~\cite{fanlv2008} to show $\lim_{n\rightarrow\infty} {\rm Pr}({\cal T}\subseteq\hat{\cal S}_\gamma)=1$ because we can treat ${\bm x}_j\circ{\bm x}_k$ as a main effect in the theoretical evaluation of our method. The approach has been previously used for the theoretical properties of the all-pair SIS by~\cite{hallxue2014}. If $\lim_{n\rightarrow\infty} {\rm Pr}({\cal T}\subseteq\hat{\cal S}_\gamma)=1$ is satisfied, then it is not necessary to consider any variables outside of $\hat{\cal S}_\gamma$ in the penalization stage because they are not related to any important main effects or interactions under the framework of the asymptotic theory. Based on the oracle properties of the SCAD and the MCP, we obtain structural oracle properties of our method if it is combined with the SCAD or the MCP. Based on the consistency of the LASSO, we obtain consistency of our method if it is combined with the LASSO. A special case is our modified \textsf{GRESH}.

We use $\lambda_{\max}(\cdot)$ and $\lambda_{\min}(\cdot)$ to denote the largest and smallest eigenvalues of a matrix, respectively. We denote ${\bm z}_{jk}={\bm x}_j\circ{\bm x}_k$ for $0\le j<k\le p$, where we use ${\bm z}_{0k}$ to represent the main effects and ${\bm z}_{jk}$ with $j,k>0$ to represent the interactions. We express the full model~\eqref{eq:definition of the full interaction effect model} as ${\bm y}=\beta_0+\sum_{j=0}^{p-1}\sum_{k=j+1}^p \beta_{jk}{\bm z}_{jk}+{\bm\epsilon}$ and the true model~\eqref{eq:true model} as ${\bm y}=\beta_0^*+\sum_{j\in {\cal T}_M}\beta_j^*{\bm z}_{0j}+\sum_{(j,k)\in{\cal T}_I}\beta_{jk}^*{\bm z}_{jk}+{\bm\epsilon}^*$.  We use ${\bf Z}_\alpha$ to denote the sub-matrix derived by choosing a subset of the columns contained by $\alpha\subseteq\{(j,k):0\le j<k\le p\}$ from the design matrix of the full model. Therefore, the columns of ${\bf Z}_\alpha$ are the main effects and interactions. Let $\alpha^*=\{(j,k): j\in{\cal T}_M~{\rm when}~k=0~{\rm or}~(j,k)\in{\cal T}_I~{\rm when}~k\not=0\}$ and ${\bm\beta}^*$ be the regression coefficients vector for the true model. Then, the true linear function is ${\rm E}({\bm y})={\bf Z}_{\alpha^*}{\bm\beta}^*$. 

We assume that the sizes of ${\cal T}$, ${\cal T}_M$, and ${\cal T}_I$, defined as $s=|{\cal T}|$, $s_M=|{\cal T}_M|$, and $s_I=|{\cal T}_I|$, respectively, also approach $\infty$ in a low order of $n$, implying that they satisfy $s=o(n)$, $s_M=o(n)$, and $s_I=o(n)$ as $n\rightarrow\infty$. We show consistency and asymptotic normality of our method under a set of regularity conditions. For consistency, we want to show $\lim_{n\rightarrow\infty}P\{\hat{\cal S}_{\gamma}\supseteq {\cal T}\}=1$ with a suitable choice of $\gamma$. For asymptotic normality, we want to show the oracle properties of the model selected by the PLS. We assume that a non-concave penalty, such as the SCAD or the MCP, is applied. We exclude the LASSO because it may induce inconsistent estimators. 

We extend the regularity conditions for the SIS for the main effects given by~\cite{fanlv2008} to propose the regularity conditions for our method. They are proposed based on the properties of a simple linear regression model as 
\begin{equation}
\label{eq:simple linear regression}
{\bm y}=\beta_0+\beta_1{\bm x}+{\bm\epsilon}, {\bm\epsilon}\sim{\cal N}(0,\sigma^2{\bf I}),
\end{equation} where ${\bm x}=(x_1,\dots,x_n)^\top$ is the unique explanatory variable. The MLE of $\beta_1$ is $\hat\beta_1=\sum_{i=1}^n(x_i-\bar x)(y_i-\bar y)/\sum_{i=1}^n (x_i-\bar x)^2=\hat\rho[\sum_{i=1}^n (x_i-\bar x)^2/\sum_{i=1}^n (y_i-\bar y)^2]^{1/2}$, where $\hat\rho={\rm cor}({\bm y},{\bm x})$ is the sample correlation coefficient between ${\bm y}$ and ${\bm x}$. The estimator of the variance of $\hat\beta_1$ is $\hat\sigma_{\hat\beta_1}^2=\widehat{\rm var}(\hat\beta_1)=\hat\sigma^2/\sum_{i=1}^n (x_i-\bar x)^2$,  where $\hat\sigma^2=[\sum_{i=1}^n (y_i-\bar y)^2-\hat\beta_1\sum_{i=1}^n (x_i-\bar x)^2]/(n-2)$ is the MSE. The $t$-value of the slope is $t_{\hat\beta_1}=\hat\beta_1/\hat\sigma_{\beta_1}=\hat\rho/[(1-\hat\rho^2)/(n-2)]^{1/2}$. For the theoretical properties of $t_{\hat\beta_1}$, we study the cases when $\beta_1^*=0$ and $\beta_1^*\not=0$, respectively, where $\beta_1^*$ represents the true value of $\beta_1$. In the first case, $t_{\hat\beta_1}\sim t_{n-2}$, implying that it is bounded in probability. In the second case, there is $t_{\beta_1}\sim t_{n-2}(\delta)$, where $\delta=\beta_1^*[\sum_{i=1}^n (x_i-\bar x)^2]^{1/2}$ is the non-centrality of the non-central $t_{n-2}$-distribution. It means that $\sum_{i=1}^n (x_i-\bar x)^2$ linearly increases with $n$ as $n\rightarrow\infty$ because it can be achieved by assuming that $x_1,\dots,x_n$ are derived by iid sampling from a certain distribution. If $\beta_1^*$ is a constant, then the magnitude of $\delta$ is proportional to $[\sum_{i=1}^n (x_i-\bar x)^2]^{1/2}$, implying that $t_{\hat\beta_1}$ is approximately proportional to $n^{1/2}$ as $n\rightarrow\infty$. We then conclude that the correlation learning proposed by~\cite{fanlv2008} can be treated as the simple regression learning based on~\eqref{eq:simple linear regression} with ${\bm x}$ specified for the corresponding main effects individually. A similar conclusion also holds in the proposed aggregated correlation learning.

\begin{lem}
\label{lem:simple aggregated regression learning}
${\rm acor}({\bm x}_j)=\max( |t_{\hat\beta_{jk}^1}|; k\in\{0,1,\dots,p\}, k\not=j)$,  where $t_{\hat\beta_{jk}^1}=\hat\beta_{jk}^1/\hat\sigma_{\hat\beta_{jk}^1}$ and $\hat\beta_{jk}^1$ is the MLE of $\beta_{jk}$ with $\hat\sigma_{\hat\beta_{jk}^1}$ being the corresponding standard error of the simple linear regression model as ${\bm y}=\beta_0+\beta_{jk}{\bm z}_{jk}+{\bm\epsilon}$,  ${\bm\epsilon}\sim{\cal N}({\bm 0},\sigma^2{\bf I})$. 
\end{lem}

Lemma~\ref{lem:simple aggregated regression learning} indicates that it is appropriate to focus on the performance of ${n}^{1/2}{\rm acor}({\bm x}_j)$ when the theoretical properties of our method is evaluated. This contains the cases when ${\bm x}_j$ is active and redundant, respectively. For each fixed $j$, if ${\bm x}_j$ is active, then the speed for ${n}^{1/2}{\rm acor}({\bm x}_j)$ to grow to $\infty$ is proportional to ${n}^{1/2}$; otherwise the speed is proportional to the absolute maximum of $p$ independent $t_{n-2}$ distributions, which is approximately equal to a constant times $[2\log{p}]^{1/2}$ when $n$ is large. To distinguish the two cases, we assume that the growth of $\log{p}$ is lower than $n^{1/2}$. If $j$ varies, then we evaluate the maximum of $p$ aggregated correlation coefficients. This does not affect the growth of $\log{p}$ as the upper bound can be adjusted to be $[4\log{p}]^{1/2}$, which is still achieved by assuming $\log{p}=o(n^{1/2})$. We propose our regularity conditions below. 

{\it Condition 1.} There exists $c_1>0$ such that $\|{\bf Z}_{\alpha^*}{\bm\beta}^*\|^2=o(n^{1+c_1})$.

{\it Condition 2.} There exists $c_2>0$ such that ${\rm cor}^2({\bm z_{jk}},{\bm y})\ge n^{-1/2+c_2}$ for any $(j,k)\in\alpha^*$.

{\it Condition 3.} There exists positive $c_3$ and $c_4$ such that $\lambda_{\min}(n^{-1}{\bf Z}_{\alpha}^\top{\bf Z}_\alpha)\ge n^{-1/2+c_3}$ and  $\lambda_{\max}(n^{-1}{\bf Z}_{\alpha}^\top{\bf Z}_\alpha)\le n^{1/2-c_4}$ for any $\alpha$ satisfying $|\alpha|\le n/\log{n}$ if $n$ is sufficiently large.  

{\it Condition 4.} (i) $p$ increases with $n$ exponentially: $p>n$ and $\log{p}=o(n^{1/2-c_5})$ for some $c_5>0$; or (ii) $p$ increases with $n$ polynomially: $p>n$ and $p=o(n^{c_5})$ for some $c_5>0$;

Note that ${\rm E}({\bm y})={\bf Z}_{\alpha^*}{\bm\beta}^*$ under the true model. Condition 1 allows the magnitude of ${\bm y}$ to increase with $n$, but the speed cannot be too fast. Condition 2 allows the absolute correction between the response and the true main or interaction effects to decrease with $n$, but the speed cannot be too fast either. Condition 3 rules out strong collinearity.  Condition 4 allows the speed of $p$ to be exponentially fast as $n$ increases, but it must be slower than $\exp(n^{1/2})$.

\begin{thm}
\label{thm:consistency of SIS} (Sure screening property with an exponentially increasing $p$). Assume that Conditions 1--3 and 4(i) are satisfied. If $c_2+c_4>1/2$, $c_1<1/2$, and $\gamma n^{1/2-c_1}\rightarrow\infty$, then $\lim_{n\rightarrow\infty}P(\hat{\cal S}_{\gamma}\supseteq{\cal T})=1$. 
\end{thm}

\begin{thm}
\label{thm:polynomial increases of p}
 (Sure screening property with a polynomially increasing $p$). 
Assume that Conditions 1--3 and 4(ii) are satisfied. If $c_1<1/2$ and $\gamma n^{1/2-c_1}\rightarrow\infty$, then $\lim_{n\rightarrow\infty}P(\hat{\cal S}_{\gamma}\supseteq{\cal T})=1$. 
\end{thm}

We point out that $\gamma{n}=o(n)$ and ${\gamma}n\rightarrow\infty$ are needed in theoretical evaluations of our method. By Theorem~\ref{thm:consistency of SIS}, if $p$ grows exponentially fast with $n$, then the growth rate of $\gamma$ cannot be very high. We need to choose more than $n^{1/2-c_1}$ variables from the candidates to ensure that $\hat{\cal S}_\gamma$ contains the true variable set, implying that we cannot make $|\hat{\cal S}_\gamma|$ very small. By Theorem~\ref{thm:polynomial increases of p}, if $p$ grows polynomially fast with $n$, then the growth rate of $\gamma$ can be very high. In the extreme case, if $c_1$ can be arbitrarily small in Condition 1 and $c_2$ and $c_4$ can be arbitrarily close to $1/2$, then the columns of ${\bf Z}_\alpha$ are almost orthogonal. It is sufficient to choose $\gamma$ satisfying $\gamma=o(n^{1-c})$ for an arbitrarily small $c$, implying that we can make $|\hat{\cal S}_\gamma|$ extremely small. 

We next evaluate the theoretical properties of the approach derived by combining the proposed AcorSIS with a variable selection method for interactions. We assume that the approach is carried out by an extension of the \textsf{GRESH} as  
\begin{equation}
\label{eq:penalized estimator}
\hat{\bm\beta}_\lambda=\mathop{\arg\!\min}_{\bm\beta}\ell_\lambda({\bm\beta}),
\end{equation}
where 
\begin{equation}
\label{eq:PLS objection function}
\eqalign{
{\mathcal L}({\bm\beta})=&{1\over 2n}\|{\bm y}-\beta_0-\sum_{j\in\hat{\cal S}_\lambda}\beta_j{\bm x}_{j}-\mathop{\sum\sum}_{j,k\in\hat{\cal S}_{\gamma},j<k}\beta_{jk}{\bm x}_{j}\circ{\bm x}_k\|^2\cr
&+\sum_{j\in\hat{\cal S}_{\gamma}} P_\lambda\left[\left(|\beta_j|^r+\sum_{k=1,k\not=j}^p |\beta_{jk}|^r\right)^{1/r}\right]+\mathop{\sum\sum}_{j,k\in\hat{\cal S}_{\gamma},j<k} P_\lambda(\beta_{jk})
}
\end{equation}
is the objective function, $P_\lambda(\cdot)$ is a penalty function, and $\lambda$ is the tuning parameter. The approach can be further specified to the SCAD or the MCP beyond.

For a given $P_\lambda(\cdot)$, the number of nonzero components of $\hat{\bm\beta}_\lambda$ is small if $\lambda$ is large. This induces a well-known problem about the determination of $\lambda$, called the tuning parameter selection problem. In the literature, the tuning parameter selection problem is addressed by the GIC approach~\cite{zhangli2010} as
\begin{equation}
\label{eq:GIC for PML}
\hat\lambda=\mathop{\arg\!\min}_{\lambda}{\rm GIC}_\kappa(\lambda),
\end{equation}
where 
\begin{equation}
\label{definition of GIC}
{\rm GIC}_\kappa(\lambda)= -2\ell(\hat\sigma_\lambda^2)+\kappa df_\lambda
\end{equation}
is the GIC objective function, $\ell(\hat{\bm\beta}_{\lambda})=-(n/2)[1+\log(2\pi)]-(n/2)\log(\hat\sigma_\lambda^2)$ with the MSE given by $\hat\sigma_\lambda^2=(1/n)\|{\bm y}-\hat\beta_{0\lambda}-\sum_{j\in\hat{\cal S}_\lambda}\hat\beta_{j\lambda}{\bm z}_{j0}-\mathop{\sum\sum}_{j,k\in\hat{\cal S}_{\gamma},j<k}\hat\beta_{jk\lambda}{\bm x}_{j}{\bm x}_{k}\|^2$ is the loglikehood function, $\hat\beta_{0\lambda}$, $\hat\beta_{j\lambda}$, and $\hat\beta_{jk\lambda}$ are the penalized estimators of $\beta_0$, $\beta_j$ and $\beta_{jk}$ given by~\eqref{eq:penalized estimator}, respectively, $df_\lambda$ is the model degrees of freedom, and $\kappa$ is a pre-selected constant that controls the properties of the GIC. Because $p$ can be exponentially fast as $n$ increases, we adopt the EBIC approach~\cite{fantang2013} by $\kappa=\log{p}\log\!\log{n}$. 

\begin{thm}
\label{thm:oracle properties}
(Structural Oracle Properties of the SCAD and the MCP). If $P_\lambda(\cdot)$ is the SCAD or the MCP in~\eqref{eq:PLS objection function} and $\kappa=\log{p}\log\!\log{n}$ in~\eqref{definition of GIC}, then $\hat{\bm\beta}_\lambda$ given by~\eqref{eq:penalized estimator} satisfies: (a) with probability one $\hat\beta_{j\lambda}=0$ iff $j\not\in{\cal T}$ and $\hat\beta_{jk\lambda}=0$ iff $(j,k)\in{\cal T}_I$; (b) the components of $\hat{\bm\beta}_\lambda$ in ${\cal T}$ and ${\cal T}_I$ perform as well as if the true model~\eqref{eq:true model} is assumed to be known.
\end{thm}

\begin{thm}
\label{thm:consistency o the LASSO}
(Consistency of the LASSO). If $P_\lambda(\cdot)$ is the LASSO in~\eqref{eq:PLS objection function} and $\kappa=\log{p}\log\!\log{n}$ in~\eqref{definition of GIC}, then $\hat{\bm\beta}_\lambda$ given by~\eqref{eq:penalized estimator} satisfies (a) of Theorem~\ref{thm:oracle properties}.
\end{thm}

\section{Simulation}
\label{sec:simulation}

We evaluate the performance of our proposed AcorSIS with the comparison to our competitors via Monte Carlo simulations with $1000$ replications. We classify our competitors into two groups. The interaction effect group is proposed for screening main effects and interactions. It contains the IPDC and the IP~\cite{kongli2017}, the SIRI~\cite{jiangliu2014}, the DCSIS~\cite{lizhong2012},  the BcorSIS~\cite{panwang2019}, and the iForm~\cite{haozhang2014}. The main effect group is proposed for screening the main effects only. It contains the SIS~\cite{fanlv2008}, the CondSIS~\cite{barutfan2016},  the MDCSIS~\cite{shaozhang2014}, the SIRS~\cite{zhuli2011}, the WLS~\cite{zhongliu2021}, the Kfilter~\cite{mai2015}, and the CSIS~\cite{chengwang2023}. We use the coverage probabilities defined as ${\rm Pr}(\hat{\cal T}\supseteq{\cal T})$ to compare these methods, where $\hat{\cal T}$ is the variable set provided by a screening method and ${\cal T}$ is the true variable set.

We assume that variable selection is investigated under the full model given by~\eqref{eq:definition of the full interaction effect model} with $n=200$ and $p=2000$. We generate rows of variable matrix ${\bf X}=[{\bm x}_1,\dots,{\bm x}_p]$ independently with dependencies between the columns by normal distributions with ${\rm cov}(x_{ij},x_{ik})=\rho^{|j-k|}$ for $i=1,\dots,n$ and $j,k=1,\dots,p$, where we choose $\rho=0.0,0.5,0.8$, respectively. After ${\bf X}$ is derived, we generate data from the true model as
\begin{equation}
\label{eq:true model in simulation}
\eqalign{
{\bm y}=&\beta_1^*{\bm x}_1+\beta_2^*{\bm x}_2+\beta_3^*{\bm x}_3+\beta_4^*{\bm x}_4+\beta_5^*{\bm x}_5+\beta_6^*{\bm x}_6\cr
&+3{\bm x}_1\circ{\bm x}_{4}+3{\bm x}_1\circ{\bm x}_{5}+3{\bm x}_5\circ{\bm x}_{6}+{\bm\epsilon}^*, {\bm\epsilon}^*\sim {\cal N}({\bm 0},{\bf I}),
}
\end{equation}
implying that we always choose $\beta_{14}^*=\beta_{15}^*=\beta_{56}^*=3$. We consider three cases of the main effects structures. Case (a) is designed for the weak hierarchy of the true model, where we choose $\beta_1^*=\beta_2^*=\beta_3^*=\beta_4^*=3$ and $\beta_5^*=\beta_6^*=0.0$, leading to ${\cal T}_M=\{1,2,3,4\}$ and ${\cal T}=\{1,2,3,4,5,6\}$. Case (b) is designed for the strong hierarchy of the true model, where we choose $\beta_1^*=\beta_2^*=\beta_3^*=\beta_4^*=\beta_5^*=\beta_6^*=3$, leading to ${\cal T}_M={\cal T}=\{1,2,3,4,5,6\}$. Case (c) is designed for no hierarchy of the true model, where we choose $\beta_1^*=\beta_2^*=\beta_3^*=\beta_4^*=\beta_5^*=\beta_6^*=0$, leading to ${\cal T}_M=\emptyset$ and ${\cal T}=\{1,4,5,6\}$. The full model contains $2000$ candidates of the main effects and $1,\!999,\!000$ candidates of the interaction effects. It is impossible to use a penalization method to select variables directly. Thus, screening is required.

\begin{table}
\scriptsize
\caption{\label{tab:simulation for coverage probabilities}Simulations with $1000$ replications for coverage probabilities of the true variable set (i.e., ${\rm Pr}(\hat{\cal T}\supseteq{\cal T})$) when data are generated from~\eqref{eq:true model in simulation}, where ${\cal T}=\{1,2,3,4,5,6\}$ in cases (a) and (b) or ${\cal T}=\{1,2,5,6\}$ in case (c).}
\begin{center}
\begin{tabular}{cccccccccc}\hline
 Screening            & \multicolumn{3}{c}{$\rho=0$} &  \multicolumn{3}{c}{$\rho=0.5$} & \multicolumn{3}{c}{$\rho=0.8$} \\\cline{2-10}
Method & (a) & (b) & (c)  & (a) & (b) & (c)  & (a) & (b) & (c) \\\hline
Proposed&$0.994$&$0.992$&$0.997$&$0.994$&$1.000$&$0.999$&$1.000$&$1.000$&$1.000$\\
IPDC&$0.003$&$0.955$&$0.002$&$0.444$&$0.999$&$0.398$&$1.000$&$1.000$&$1.000$\\
IP&$0.001$&$0.914$&$0.000$&$0.309$&$1.000$&$0.012$&$1.000$&$1.000$&$0.096$\\
SIRS&$0.064$&$0.772$&$0.344$&$0.013$&$0.146$&$0.514$&$0.000$&$0.000$&$0.062$\\
DCSIS&$0.002$&$0.899$&$0.000$&$0.288$&$1.000$&$0.015$&$0.994$&$1.000$&$0.100$\\
BcorSIS&$0.001$&$0.925$&$0.000$&$0.113$&$1.000$&$0.001$&$0.986$&$1.000$&$0.011$\\
iForm&$0.004$&$0.002$&$0.959$&$0.069$&$0.082$&$0.795$&$0.990$&$1.000$&$0.969$\\
SIS&$0.003$&$0.961$&$0.003$&$0.475$&$1.000$&$0.561$&$1.000$&$1.000$&$1.000$\\
CondSIS&$0.001$&$0.768$&$0.000$&$0.002$&$0.122$&$0.957$&$0.000$&$0.029$&$1.000$\\
MDCSIS&$0.026$&$0.684$&$0.164$&$0.359$&$0.999$&$0.854$&$1.000$&$1.000$&$1.000$\\
SIRI&$0.000$&$0.000$&$0.000$&$0.000$&$0.000$&$0.000$&$0.000$&$0.000$&$0.000$\\
WLS&$0.001$&$0.072$&$0.000$&$0.030$&$0.960$&$0.000$&$0.949$&$1.000$&$0.090$\\
Kfilter&$0.001$&$0.913$&$0.000$&$0.089$&$1.000$&$0.001$&$0.988$&$1.000$&$0.013$\\
CSIS&$0.005$&$0.999$&$0.000$&$0.001$&$0.992$&$0.000$&$0.003$&$0.945$&$0.000$\\\hline
\end{tabular}
\end{center}
\end{table}

We assumed that the variable selection procedure has two stages. The first was the screening stage which was our primary interest. The goal was to provide a variable set that contains ${\mathcal T}$. The second was the penalization stage which was our minor interest. The goal was to select the main effects and the interactions from the variable set provided by the screening stage. In the first stage, we implemented our proposed method with comparisons to our competitors. The simulations showed that our proposed method could identify the true variable set in all the cases we studied (Table~\ref{tab:simulation for coverage probabilities}). If columns were independent and the strong hierarchical restriction was violated in the true model (i.e., (a) and (c) when $\rho=0$), then our competitors rarely identified the true variable sets. The situation was better if the strong hierarchical restriction was satisfied in the true model (i.e., (b) when $\rho=0$). The probabilities for our competitors to correctly identify the true variable sets generally increased with $\rho$. This could be interpreted because ${\rm cor}(x_{ij},x_{ik})$ increased with $\rho$ when $|j-k|$ was small. 

\begin{table}
\caption{\label{tab:simulation for true positives} Simulations with $1000$ replications for the proportions of the true positives and the numbers of false positives of the main (M) and interaction (I) effects when the proposed AcorSIS method is combined with the previous \textsf{GRESH}, \textsf{SHIM}, and \textsf{hierNet} methods, where the all-pairs SIS is implemented in the previous \textsf{HiQR} and \textsf{sprinter} methods, a two-stage regularization SIS is implemented in the previous \textsf{RAMP} method, and a proximal SIS is implemented in the previous \textsf{hierScale} method.}
\begin{center}
\begin{tabular}{ccccccccccc}\hline
  &  &\multicolumn{3}{c}{$\rho=0.0$}& \multicolumn{3}{c}{$\rho=0.5$} & \multicolumn{3}{c}{$\rho=0.8$} \\\cline{3-11}
Method  & Effect &  (a) & (b) & (c) & (a) & (b) & (c)& (a) & (b) & (c)  \\\hline
 &  & \multicolumn{9}{c}{Proportions of True Positives}\\
\textsf{GRESH$^*$}  & M  &   $0.989$ & $0.953$ & $0.999$ & $0.998$ & $0.999$ & $0.999$ & $1.000$ & $1.000$ & $1.000$   \\
  & I  & $0.993$ & $0.961$ & $0.999$ & $0.998$ & $0.999$ & $0.999$ & $1.000$ & $1.000$ & $1.000$\\
\textsf{SHIM$^*$} &  M & $0.890$ & $0.957$ & $0.557$ & $0.912$ & $0.999$ & $0.782$ & $0.860$ & $0.997$ & $0.946$  \\
   & I  & $0.797$ & $0.967$ & $0.552$ & $0.809$ & $0.999$ & $0.681$ & $0.161$ & $0.997$ & $0.898$ \\
\textsf{hierNet$^*$} & M & $0.000$ & $0.000$ & $0.306$ & $0.000$ & $0.000$ & $0.015$ & $0.003$ & $0.000$ & $0.071$ \\
& I &   $0.000$ & $0.000$ & $0.306$ & $0.000$ & $0.000$ & $0.015$ & $0.003$ & $0.000$ & $0.071$  \\
\textsf{HiQR} &  M &  $0.658$ & $0.956$ & $0.000$ & $0.670$ & $0.999$ & $0.000$ & $0.684$ & $1.000$ & $0.000$ \\
 & I  & $0.990$ & $0.967$ & $0.998$ & $0.998$ & $0.999$ & $0.998$ & $1.000$ & $1.000$ & $1.000$ \\
\textsf{springer} & M  &  $0.742$ & $0.920$ & $0.065$ & $0.729$ & $0.999$ & $0.657$ & $0.737$ & $1.000$ & $0.262$  \\
 & I & $0.950$ & $0.283$ & $0.336$ & $0.997$ & $0.997$ & $0.852$ & $1.000$ & $1.000$ & $1.000$\\
\textsf{RAMP} & M  &  $0.676$&$0.999$&$0.024$&$0.680$&$1.000$&$0.005$&$0.440$&$0.861$&$0.057$ \\
& I & $0.347$&$0.995$&$0.002$&$0.355$&$1.000$&$0.002$&$0.119$&$0.704$&$0.004$ \\
\textsf{hierScale} & M & $0.609$ & $0.919$ & $0.600$ & $0.855$ & $1.000$ & $0.911$ & $0.967$ & $1.000$ & $1.000$  \\
  & I &   $0.116$ & $0.742$ & $0.533$ & $0.684$ & $0.999$ & $0.863$ & $0.935$ & $1.000$ & $1.000$ \\\hline
  &   &  \multicolumn{9}{c}{Numbers of False Positives}\\
\textsf{GRESH$^*$} & M &  $7.661$ & $6.848$ & $8.764$ & $7.393$ & $6.593$ & $8.484$ & $5.635$ & $4.517$ & $7.925$    \\
& I & $3.593$ & $2.392$ & $4.764$ & $7.678$ & $6.159$ & $7.836$ & $12.885$ & $10.387$ & $14.759$ \\
\textsf{SHIM$^*$} &  M &  $2.555$ & $0.140$ & $4.741$ & $1.999$ & $0.048$ & $6.063$ & $6.113$ & $0.508$ & $4.701$ \\
& I & $2.390$ & $6.930$ & $0.588$ & $6.233$ & $6.093$ & $0.632$ & $1.107$ & $5.833$ & $2.705$ \\
\textsf{herNet$^*$} &  M & $0.872$ & $0.673$ & $7.885$ & $0.969$ & $0.274$ & $2.969$ & $1.113$ & $0.167$ & $2.007$  \\
 & I & $1.000$ & $0.060$ & $4.477$ & $1.000$ & $1.001$ & $2.684$ & $1.081$ & $1.108$ & $3.859$ \\
\textsf{HiQR} & M  & $0.025$ & $0.060$ & $1.000$ & $0.007$ & $0.026$ & $1.001$ & $0.011$ & $0.115$ & $0.999$   \\
& I & $5.315$ & $8.630$ & $1.648$ & $2.745$ & $3.805$ & $1.038$ & $2.419$ & $2.626$ & $1.679$ \\
\textsf{sprinter} & M & $0.120$ & $0.250$ & $0.871$ & $0.037$ & $0.082$ & $0.614$ & $0.065$ & $0.173$ & $0.638$ \\
 & I & $5.110$ & $8.070$ & $3.021$ & $2.179$ & $2.826$ & $1.340$ & $2.295$ & $2.383$ & $1.901$ \\
\textsf{RAMP} & M  & $0.287$&$0.114$&$1.004$&$0.547$&$0.085$&$0.997$&$0.577$&$1.370$&$1.376$\\
& I &  $0.944$&$0.067$&$1.024$&$0.095$&$0.001$&$0.995$&$1.943$&$2.907$&$1.197$\\
\textsf{hierScale} & M & $0.387$ & $3.736$ & $7.945$ & $2.058$ & $0.938$ & $5.870$ & $0.601$ & $0.101$ & $3.373$ \\
  & I &  $0.001$&$0.007$&$0.145$ &$0.530$&$0.052$&$0.990$&$4.947$&$1.583$ & $8.153$ \\\hline 
\end{tabular}
\end{center}
\end{table}

We next evaluated the performance of our method when it was combined with the previous \textsf{GRESH}, \textsf{SHIM}, and \textsf{hierNet} methods for selecting the main effects and interactions. To compare, we also evaluated the built-in SIS in the previous \textsf{HiQR}, \textsf{sprinter}, \textsf{RAMP}, and \textsf{hierScale} methods. The built-in SIS adopted by the \textsf{HiQR} and \textsf{sprinter} is the all-pairs SIS described at the end of Section~\ref{subsec:aggregated correlation learning for screening Interactions}. A concern is that the all-pairs SIS does not consider the SH restriction. The built-in SIS adopted by the \textsf{RAMP} wants to capture the important main effects first and then the interactions. A concern is that it is unlikely to capture an important interaction if its related main effects are weak or absent. The built-in SIS adopted by the \textsf{hierScale} is developed based on the proximal gradient descent (PGD). The criterion is based on a summation but not a maximization, leading to a similar concern in the \textsf{RAMP}. 

The previous \textsf{GRESH}, the \textsf{SHIM}, and the \textsf{hierNet} cannot be used for the full model. Our simulation showed that without screening when $p=500$ each computation of the \textsf{GRESH} and the \textsf{SHIM} took more than $7$ days and each of the \textsf{hierNet} took more than $3$ days. With the AcorSIS, the computational time was less than $10$ seconds. We then decided to combine the AcorSIS with the three methods, leading to the modified \textsf{GRESH}, \textsf{SHIM}, and \textsf{hierNet} methods, respectively. We directly implemented the recent \textsf{HiQR}, \textsf{sprinter}, \textsf{RAMP}, and \textsf{hierScale} methods without using our proposed AcorSIS. 

 Following the traditional approach for the comparison, we calculated the numbers of true and false positives for the main and interaction effects for the five methods, respectively. The true and false positive sets for the main effects were computed by $TP_M=\{j:\hat\beta_j\not=0,\beta_j^*\not=0\}$ and $FP_M=\{j:\hat\beta_j\not=0,\beta_j^*=0\}$, respectively. The true and false positive sets for the interaction effects were computed by $TP_I=\{(j,k):\hat\beta_{jk}\not=0,\beta_{jk}^*\not=0\}$ and $FP_I=\{(j,k):\hat\beta_{jk}\not=0,\beta_{jk}^*=0\}$, respectively. After they were derived, we computed the proportions of true positives of the main effects by $|TP_M|/6$ in Cases (a) and (b) and $|TP_M|/4$ in Case (c), respectively, and the proportions of true positives for the interaction effects by $|TP_I|/3$ in all the three cases (Table~\ref{tab:simulation for true positives}, the first half). We also computed the number of false positives of the main and the interactions by $|FP_M|$ and $|FP_I|$, respectively (Table~\ref{tab:simulation for true positives}, the second half). Our modified \textsf{GRESH} method could capture all of the important main effects and interactions in general. Our modified \textsf{SHIM} could capture important interaction effects if their related main effects were important but not otherwise. Our modified \textsf{hierNet} missed most of the important main effects and interactions. It was unlikely to use the recent \textsf{HiQR} and \textsf{sprinter} to capture the active weak main effects even if they were related to important interaction effects. It was hard to use the recent \textsf{RAMP} and \textsf{hierScale} to capture important interactions if their related main effects were weak or absent because they missed ${\bm x}_5\circ{\bm x}_6$ in Cases (a) and (c) when $\rho=0$. 

\begin{table}
\caption{\label{tab:violations of strong hierarchical restrictions} Simulations with $1000$ replications for the proportions of satisfications of the SH restriction.}
\begin{center}
\begin{tabular}{ccccccccccc}\hline
  & \multicolumn{3}{c}{$\rho=0.0$}& \multicolumn{3}{c}{$\rho=0.5$} & \multicolumn{3}{c}{$\rho=0.8$} \\\cline{2-10}
  & (a) & (b) & (c) & (a) & (b) & (c)& (a) & (b) & (c)  \\\hline
\textsf{GRESH}$^*$ & $1.000$ &$1.000$ & $1.000$ &$1.000$ &$1.000$ &$1.000$ &$1.000$ &$1.000$& $1.000$ \\
\textsf{SHIM}$^*$ & $1.000$ &$1.000$ & $1.000$ &$1.000$ &$1.000$ &$1.000$ &$1.000$ &$1.000$& $1.000$   \\
\textsf{hierNet}$^*$ & $1.000$ &$1.000$ & $1.000$ &$1.000$ &$1.000$ &$1.000$ &$1.000$ &$1.000$& $1.000$  \\
\textsf{HiQR} & $0.000$ & $0.000$& $0.000$ &  $0.000$  & $0.170$& $0.000$ & $0.001$  & $0.590$ & $0.000$  \\
\textsf{springer} & $0.019$ & $0.019$ & $0.019$ & $0.014$ & $0.250$ & $0.001$ & $0.052$ & $0.748$ &  $0.001$\\
\textsf{RAMP} &  $1.000$ &$1.000$ & $1.000$ &$1.000$ &$1.000$ &$1.000$ &$1.000$ &$1.000$& $1.000$  \\
\textsf{hierScale} & $1.000$ &$1.000$ & $1.000$ &$1.000$ &$1.000$ &$1.000$ &$1.000$ &$1.000$& $1.000$  \\\hline
\end{tabular}
\end{center}
\end{table}

We next studied the satisfaction of the SH restriction. We found that it was violated by the recent \textsf{HiQR} and the \textsf{sprinter} because they used all-pairs SIS for screening the main effects and the interactions (Table~\ref{tab:violations of strong hierarchical restrictions}).  In addition, we evaluated the \textsf{SODA} based on a logistic model modified from~\eqref{eq:true model in simulation}. We found that it violated the SH restriction either (the details were not shown). This means the three methods should be treated as inappropriate in selecting the main effects and the interactions because the SH restriction is required according to the statistical literature. 

In addition, we studied the case when $p=10^4$. We found that the performance of our method did not change too much. The only difference was the time duration. The implementation of the AcorSIS took about $0.2$ minutes when $p=2000$ and $4.9$ minutes when $p=10^4$. The computation of the iForm took about $3.6$ minutes when $p=2000$ and $45.5$ minutes when $p=10^4$. The computations of the other competitors were more computationally efficient than the iForm because they were formulated based on the main effect screening technique. When $p=10^4$, the full model has $49,\!995,\!000$ candidates of interactions. This means that our method is computationally efficient even if $p$ is extremely large. 

In summary, we find that only our method can identify the true variable set no matter whether the strong hierarchical restriction is satisfied in the true model or not. In the screening stage, it is important to contain all active variables with a high probability. The goal is to ensure that the following penalization stage can find all the important main effects and interactions. The screening stage is more important than the penalization stage in variable selection for interactions.

\section{Application}
\label{sec:application}

We applied our method to the Prostate Cancer dataset. The dataset was originally provided by~\cite{singh2002} and is available in the package \textsf{SIS} of \textsf{R}. It was a gene expression microarray dataset containing $12600$ genes with $77$ prostate cancer patients and $59$ normal specimens. Prostate cancer is one of the four most common cancers in the United States. It had about $250$ thousand new cases and $34$ thousand new deaths in $2023$. The adoption of the prostate cancer screening system has led to a high probability of earlier detection, giving opportunities to cure the cancer by surgery. A challenge is to identify patients at risk for relapse with a better understanding of the molecular abnormalities of tumors at risk for relapse. The usage of microarray data can address the corresponding challenges.

The Prostate Cancer dataset has $136$ rows and $12,\!601$ columns. The response ${\bm y}$ is a $136$-dimensional vector for a Bernoulli response with $y_i=1$ if the $i$th person is a prostate cancer patient or $y_i=0$ otherwise, $i\in{\cal D}=\{1,\dots,136\}$. The size of the feature matrix ${\bf X}$ is $136\times 12600$, implying that there were $n=136$ and $p=12,\!600$ in our method. The full model was
\begin{equation}
\label{eq:logistic full model}
\log{\pi_i\over1-\pi_i}=\beta_0+\sum_{j=1}^{12600}\beta_j{\bm x}_j+\sum_{j=1}^{12599}\sum_{k=j+1}^{12600}\beta_{jk}{\bm x}_j\circ{\bm x}_k
\end{equation}
where $\pi_i={\rm Pr}(y_i=1)$. The full model has $12,\!600$ main effects and $79,\!373,\!700$ interactions. If the penalized maximum likelihood (PML) method were directly used, then the computer would allocate $75{\rm GB}$ memory in size to store the design matrix  of~\eqref{eq:logistic full model}. Using the proposed AcorSIS method, we reduced the memory needed to $15{\rm MB}$ in size, implying that variable selection can be implemented.

We used~\eqref{eq:definition of correlation for 0 or 1} to compute ${\rm acor}({\bm x}_j)$ for each $j=1,\dots,12600$. We screened the largest $25$ variables, meaning that we chose $[\gamma n]=[136/\log{136}]=25$ in the derivation of $\hat{\cal S}_{\gamma}$. We then carried out a PML method to select variables. It reported four potential interactions. We checked the significance of each. We found only one was important. We obtained the final selected model as 
\begin{equation}
\label{eq:final selected model strong}
\eqalign{
\log{\pi_i\over 1-\pi_i}=&3.11+2.31(10^{-3}){\bm x}_{4544}-4.60(10^{-2})x_{6185}\cr
&+7.72(10^{-5}){\bm x}_{4544}\circ{\bm x}_{6185}.\cr
}\end{equation}
 The $p$-values for ${\bm x}_{4544}$, $x_{6185}$, and ${\bm x}_{4544}\circ{\bm x}_{6185}$ were $0.386$, $1.55(10^{-6})$, and $7.39(10^{-6})$, respectively. The main effect ${\bm x}_{4544}$ was not significant. The residual deviance of the null model (i.e., the intercept-only model) was $G_{null}^2=186.1$. The residual deviance of the selected model was $G_{res}^2=69.1$. It provided a better model fit.

If the SH restriction were ignored, then the main effect ${\bm x}_{4545}$ would not be selected. The selected model would contain the main effect $x_{6185}$ and the interaction ${\bm x}_{4544}\circ{\bm x}_{6185}$. Consideration of the SH restriction affected the format of the final selected model. Because the SH restriction should be treated as the default, we treated~\eqref{eq:final selected model strong} as the final model in our method.


To compare, we also studied our competitors. We found that none of them captured ${\bm x}_{4544}$ in their screening stage, implying that they could not catch ${\bm x}_{4544}\circ {\bm x}_{6185}$ in their final selected models. Among those, the \textsf{iForm} took around $50$ minutes. It only found the main effect ${\bm x}_{11052}$ with the residual deviance of the corresponding selected model as $G_{res}^2=151.2$. The \textsf{RAMP} did not find any main effects or interactions, leading to the null model as its final selected model. The \textsf{SODA} did not report any main effects or interactions. It also reported the null model as its final selected model. The implementation of the \textsf{hierScale} failed because it could be used to logistic models. For the \textsf{hierNet} package of \textsf{R}, we obtained an error message saying that it failed to allocate a vector of size $80.4{\rm GB}$. For the \textsf{BOLTSSIRR} package of \textsf{R}, we obtained a report with $121$ main effects and $136$ interaction effects. We treated the result as inappropriate. For the previous screening methods, the IPDC took $1.17$ minutes. It screened individual main effects of ${\bm x}_j^2$ with the response modified as ${\bm y}^2$. The IPDC captured ${\bm x}_{6815}$ with the best model containing main effects ${\bm x}_{6643}$ and ${\bm x}_{12148}$ and the corresponding $G_{res}^2=119.6$. The SIRI captured neither ${\bm x}_{4544}$ nor ${\bm x}_{6185}$. The best model under the SIRI had main effects ${\bm x}_{3652}$ and ${\bm x}_{11818}$ and interaction effect ${\bm x}_{3652}\circ{\bm x}_{11818}$ with the corresponding $G_{res}^2=123.5$. The DCSIS took $48$ seconds and reported main effects ${\bm x}_{6643}$ and ${\bm x}_{8768}$ with no interactions and the corresponding $G_{res}^2=117.30$. The WLS took $3$ seconds and reported main effects ${\bm x}_{205}$ and ${\bm x}_{12448}$ and no interactions with the corresponding $G_{res}^2=112.19$. The CSIS took $21$ seconds and reported main effects ${\bm x}_{6185}$ and ${\bm x}_{10537}$ and no interactions with the corresponding $G_{res}^2=96.50$. The SIS, SIRS, and MDCSIS took less than $4$ seconds. They did not report any main effects or interactions. The BcorSIS took $7$ seconds and reported main effects ${\bm x}_{8850}$ and ${\bm x}_{10494}$ and their interaction with the corresponding $G_{res}^2=86.25$. The Kfilter took $6$ seconds and reported main effects ${\bm x}_{2578}$ and ${\bm x}_{9850}$ and their interaction with the corresponding $G_{res}^2=123.4$. 

We next used the validation approach with $1000$ replications to compare the performance of the final selected models reported by our method and its competitors if they did not report the null model as their final selected model. With $50\%$, we independently assigned an observation to a training dataset denoted by ${\cal D}_{train}$ or a testing dataset denoted by ${\cal D}_{test}$ satisfying ${\cal D}_{train}\cup{\cal D}_{test}={\cal D}$ and ${\cal D}_{train}\cap{\cal D}_{test}=\emptyset$. We estimated the regression coefficient vector based on the training data to the final selected models reported by our method or its competitors. We used the estimated regression coefficient vector to compute the predicted probability of the observations contained in the testing data. We then used the prediction deviance defined as 
\begin{equation}
\label{eq:deviance of the testing data}
G_{test}^2=2\sum_{i\in {\cal D}_{test}}\{y_i\log(y_i/\hat y_i)+(1-y_i)\log[(1-y_i)/(1-\hat y_i)]\},
\end{equation}
where $\hat y_i$ was the predicted value of $y_i$ for $i\in{\cal D}_{test}$, to compare the final selected models reported by our method and its competitors. We obtained $G_{test}^2=43.34$ by our method, $G_{test}^2=78.47$ by the \textsf{iForm}, $G_{test}^2=69.94$ by the IPDC, $G_{test}^2=70.10$ by the SIRI, and $G_{test}^2=69.88$ by the DCSIS, $G_{test}^2=60.78$ by the WLS, $G_{test}^2=53.25$ by the CSIS,   $G_{test}^2=56.19$ by the BcorSIS, and  $G_{test}^2=69.88$ by the Kfilter. The validation study supported that our method was the best. 

In summary, the application shows that our method is computationally efficient. It can also screen interaction structures for logistic linear models even when $p>10^4$. Our method can capture important interaction effects even if the related main effects are weak or absent. It is hard to achieve by our competitors. 

\section{Discussion}
\label{sec:discussion}

In this article, we study the challenging problem of selection of interactions for ultrahigh-dimensional data. We do not require the SH restriction to be satisfied in the true model. We need it to be satisfied in the selected model. We study three scenarios in the true model. The first scenario assumes that the true model satisfies the SH restriction. It requires both the related main effects to be important if an interaction effect is important. The second scenario assumes that the true model satisfies the weak hierarchy restriction. It requires at least one of the related main effects to be important. The third scenario does not assume any restriction in the true model. It allows both the related main effects to be unimportant if an interaction is important. Our research shows that only our method can address the third scenario, implying that it is more appropriate than our competitors. 

A two-stage procedure is needed for selections of interactions when $p$ is large. Our research shows that the screening stage is more critical than the penalization stage. No matter which kind of hierarchy restriction is needed in the true model, a strong hierarchy must be adopted in the screening stage.  Based on this, we conclude that SIS methods for active variables and penalization methods for selecting main effects and interactions can be developed separately, implying that penalization is flexible in our method.

Our idea based on the aggregated correlation coefficient can be easily extended to screening interaction effects in HDGLMs for exponential family distribution if the likelihood ratio statistic is used. We expect that the resulting method is computationally efficient with the memory needed to be $O(np)$. In addition, our idea can be extended to screening higher-order interaction effects. This is left to future research.

\appendix

\section{Coordinate Descent for The GRESH}
\label{sec:coordinate descent or the gresh}

For the \textsf{GRESH} proposed by~\cite{she2018}, it is more efficient to use the coordinate descent rather than the ADMM to optimize the objective function given by~\eqref{eq:GRESH Lq} with $r=2$, denoted as ${\mathcal L}_{\lambda_1,\lambda_2}^{GRESH}({\bm\beta})= {\mathcal L}_{\lambda_1,\lambda_2,2}^{GRESH}({\bm\beta})$. For $\beta_j>0$, if $\|{\bm\xi}_j\|>0$ with ${\bm\xi}_j=(\beta_{j1},\dots,\beta_{j(j-1)},\beta_{j(j+1)},\dots,\beta_{jp})^\top$, then
$$
{\partial{\mathcal L}_{\lambda_1,\lambda_2}^{GRESH}({\bm\beta})\over \partial\beta_j}= -{\bm x}_j^\top(\check{\bm y}-{\bm x}_j\beta_j)+n\lambda_2{\rm sign}(\beta_j), 
$$
leading to
$$\eqalign{
\check \beta_j=\left\{ \begin{array}{ll} 
0, &  {\bm x}_j^\top\check{\bm y}- n\lambda_2  \le 0 \le  {\bm x}_j^\top\check{\bm y}+n\lambda_2 ,\cr
 {{\bm x}_j^\top\check{\bm y}+n\lambda_2\over \|{\bm x}_j\|^2}, &   {\bm x}_j^\top\check{\bm y}+n\lambda_2< 0, \cr 
 {{\bm x}_j^\top\check{\bm y}-n\lambda_2\over \|{\bm x}_j\|^2}, &  {\bm x}_j^\top\check{\bm y} -  n\lambda_2> 0  .\cr \end{array}    \right. 
}$$
If $\|{\bm\xi}_j\|>0$, then
$$
{\partial{\mathcal L}_{\lambda_1,\lambda_2}^{GRESH}({\bm\beta})\over \partial\beta_j}=-{\bm x}_{j}^\top(\check{\bm y}-{\bm x}_{j}\beta_{j})+{n\lambda_2\beta_j \over (\beta_j^2+\|{\bm\xi}_j\|^2)^{1/2} }
$$
and
$$
{\partial^2{\mathcal L}_{\lambda_1,\lambda_2}^{GRESH}({\bm\beta})\over \partial\beta_j^2}=\|{\bm x}_{j}\|^2+{n\lambda_2\|{\bm\xi}_j\|^2\over (\beta_j^2+\|{\bm\xi}_j\|^2)^{3/2}}.
$$
We devise a Newton-Raphson algorithm to compute the optimizer of the objective function with the solution denoted as $\check\beta_j$. In both cases, we can compute $\check\beta_j$. 

Let ${\bm\xi}_{j}^k$ be the vector derived by removing $\beta_{jk}$ from $(\beta_j,{\bm\xi}_j^\top)^\top$. For $\beta_{jk}$, if $\|{\bm\xi}_j^k\|=\|{\bm\xi}_k^j\|=0$, then 
$$
{\partial{\mathcal L}_{\lambda_1,\lambda_2}^{GRESH}({\bm\beta})\over \partial\beta_{jk}}= -{\bm x}_{jk}^\top(\check{\bm y}-{\bm x}_{jk}\beta_{jk})+2n\lambda_2{\rm sign}(\beta_{jk})+n\lambda_1{\rm sign}(\beta_{jk}),
$$
leading to
$$
\check\beta_{jk}=\left\{\begin{array}{ll} 0,&    {\bm x}_{jk}^\top\check y-n(2\lambda_2+\lambda_1)\le 0 \le   {\bm x}_{jk}^\top\check y+n(2\lambda_2+\lambda_1), \cr {{\bm x}_{jk}^\top\check y+n(2\lambda_2+\lambda_1)\over\|{\bm x}_{jk}\|^2}, &  {\bm x}_{jk}^\top\check y+n(2\lambda_2+\lambda_1)<0,\cr  {{\bm x}_{jk}^\top\check y-n(2\lambda_2+\lambda_1)\over\|{\bm x}_{jk}\|^2}, &  {\bm x}_{jk}^\top\check y-n(2\lambda_2+\lambda_1)>0.\cr    \end{array}    \right.
$$
If $\|{\bm\xi}_j^k\|=0$ and $\|{\bm\xi}_k^j\|>0$, then 
$$\eqalign{
{\partial{\mathcal L}_{\lambda_1,\lambda_2}^{GRESH}({\bm\beta})\over \partial\beta_{jk}}=& -{\bm x}_{jk}^\top(\check{\bm y}-{\bm x}_{jk}\beta_{jk}) +n(\lambda_2+\lambda_1){\rm sign}(\beta_{jk}) +{n\lambda_2\beta_{jk}\over(\beta_{jk}^2+\|{\bm\xi}_k^j\|^2)^{1/2}}
}$$
and
$$
{\partial^2{\mathcal L}_{\lambda_1,\lambda_2}^{GRESH}({\bm\beta})\over \partial\beta_{jk}^2}= \| {\bm x}_{jk}\|^2+{n\lambda_2\|{\bm\xi}_k^j\|^2\over(\beta_{jk}^2+\|{\bm\xi}_k^j\|^2)^{3/2}}.
$$
We then devise a Newton-Raphson algorithm to optimize the objective function for cases when $\beta_{jk}<0$ and $\beta_{jk}>0$, respectively. Let the solution be denoted by $\check\beta_{jk}$. We show that at most one solution matches the domain. If neither matches, then we set $\check\beta_{jk}=0$. Similarly, we can work out the case when $\|{\bm\xi}_j^k\|>0$ and $\|{\bm\xi}_k^j\|=0$. If $\|{\bm\xi}_j^k\|>0$ and $\|{\bm\xi}_k^j\|>0$, then 
$$\eqalign{
{\partial{\mathcal L}_{\lambda_1,\lambda_2}^{GRESH}({\bm\beta})\over \partial\beta_{jk}}=&  -{\bm x}_{jk}^\top(\check{\bm y}-{\bm x}_{jk}\beta_{jk}) +n\lambda_1{\rm sign}(\beta_{jk})\cr
 &+{n\lambda_2\beta_{jk}\over(\beta_{jk}^2+\|{\bm\xi}_j^k\|^2)^{1/2}}+{n\lambda_2\beta_{jk}\over(\beta_{jk}^2+\|{\bm\xi}_k^j\|^2)^{1/2}}\cr
}$$
and
$$
{\partial^2{\mathcal L}_{\lambda_1,\lambda_2}^{GRESH}({\bm\beta})\over \partial\beta_{jk}^2}= \| {\bm x}_{jk}\|^2+{n\lambda_2\|{\bm\xi}_j^k\|^2\over(\beta_{jk}^2+\|{\bm\xi}_j^k\|^2)^{3/2}}+{n\lambda_2\|{\bm\xi}_k^j\|^2\over(\beta_{jk}^2+\|{\bm\xi}_k^j\|^2)^{3/2}}.
$$
Based on the above, we devise a Newton-Raphson algorithm for $\check\beta_{jk}$. We obtain a coordinate descent algorithm for the \textsf{GRESH} when $r=2$.

\section{Proofs}
\label{sec:proofs}

\noindent
{\bf Proof of Lemma~\ref{lem:simple aggregated regression learning}.}  The conclusion can be easily shown by the connection between the sample correlation coefficient and the $t$-value for the slope of the simple linear regression. \qed

\noindent
{\bf Proof of Theorem~\ref{thm:consistency of SIS}.}  Without the loss of generality, we assume that the response, the main effects, and the interactions are standardized because standardization does not change the correlation matrix. It is enough to focus on the case when ${\bm 1}^\top{\bm y}={\bm 1}^\top{\bm z}_{jk}=0$ and $\|{\bm y}\|^2=\|{\bm z}_{jk}\|^2=n$. In this setting, we have ${\rm cor}({\bm y},{\bm z}_{jk})={\bm z}_{jk}^\top{\bm y}/n$ and ${\rm acor}({\bm x}_j)=\max_{0\le k\le p,k\not=j}|{\rm cor}({\bm y},{\bm z}_{jk})|$. For any $\alpha$ satisfying $|\alpha|< n/\log{n}$, the MLE of the model ${\bm y}={\bf Z}_\alpha{\bm\beta}_\alpha+{\bm\epsilon}$, ${\bm\epsilon}\sim {\cal N}({\bm 0},\sigma^2{\bf I})$, is $\hat{\bm\beta}_\alpha=({\bf Z}_{\alpha}^\top{\bf Z}_{\alpha})^{-1}{\bf Z}_\alpha{\bm y}$. By $\|{\rm E}({\bf Z}_\alpha\hat{\bm\beta}_\alpha)\|=\|{\bf Z}_\alpha({\bf Z}_{\alpha}^\top{\bf Z}_{\alpha})^{-1}{\bf Z}_\alpha^\top{\bf Z}^*{\bm\beta}^*\|\le \|{\bf Z}^*{\bm\beta}^*\|$, we obtain $ \|{\bf Z}_\alpha\hat{\bm\beta}_\alpha\|^2\le 2\{ \| {\rm E}({\bf Z}_\alpha\hat{\bm\beta}_\alpha)\|^2+\|{\bf Z}_\alpha\hat{\bm\beta}_\alpha -{\rm E}({\bf Z}_\alpha\hat{\bm\beta}_\alpha)\|^2\}\le 2 \|{\bf Z}^*{\bm\beta}^{*}\|^2+2\|{\bf Z}_\alpha\hat{\bm\beta}_\alpha -{\rm E}({\bf Z}_\alpha\hat{\bm\beta}_\alpha)\|^2$. By $\|{\bf Z}_\alpha\hat{\bm\beta}_\alpha -{\rm E}({\bf Z}_\alpha\hat{\bm\beta}_\alpha)\|^2\sim \sigma^2\chi_{|\alpha|}^2$, for $K=[\gamma{n}]$, we have
$$\eqalign{
&{\rm Pr}\{\max_{|\alpha|=K}K^{-1}\|{\bf Z}_\alpha\hat{\bm\beta}_\alpha\|^2>2t +{2\over K}\|{\bf Z}^*{\bm\beta}^*\|^2\}\cr
 \le &\sum_{|\alpha|=K}{\rm Pr}\{K^{-1}\|{\bf Z}_\alpha\hat{\bm\beta}_\alpha\|^2>2t +{2\over K}\|{\bf Z}^*{\bm\beta}^*\|^2\}\cr
\le &\sum_{|\alpha|=K}{\rm Pr}\{\|{\bf Z}_\alpha\hat{\bm\beta}_\alpha-{\rm E}({\bf Z}_\alpha\hat{\bm\beta}_\alpha) \|^2>Kt \}\cr
\le & p^K {\rm Pr}\{\chi_{K}^2\ge Kt\}\cr
\le & {Kt\over\sqrt{\pi}(Kt-K+2)}\exp\left\{ -{1\over 2}\left[ Kt- K+(K-2)\log{t}+\log{K}    \right] + K\log{p}  \right\},
}$$
which approaches $0$ if $t/n^{1/2}\rightarrow\infty$ as $n\rightarrow\infty$ by Condition 4(i). The derivation of the above needs the upper tail probability of the $\chi^2$-distribution given by~\cite{inglot2010} as
$$
{\rm Pr}\{\chi_r^2\ge u\}\le {u\over\sqrt{\pi}(u-r+2)}\exp\left\{-{1\over 2}[u-r-(r-2)\log(u/r)+\log{r}]\right\},
$$
for any $r\ge 2$ and $u\ge r-2$. By Condition 1, we have $(2/K)\|{\bf Z}^*{\bm\beta}\|^2=o(n^{c_1}/\gamma)=o(n^{1/2})$. Thus, we can ignore $(2/K)\|{\bf Z}^*{\bm\beta}\|^2$ in the evaluation of the asymptotic properties of the inequality if we choose $t/n^{1/2}\rightarrow\infty$. We obtain $\max_{|\alpha|=K}K^{-1}\|{\bf Z}_\alpha\hat{\bm\beta}_\alpha\|^2=o(n^{1/2+c})$ for any $c>0$. We next evaluate the properties of $\max_{|\alpha|=[\gamma n]}\|{\bf Z}_\alpha\hat{\bm\beta}_\alpha\|^2$ as $n\rightarrow\infty$. Using Condition 3, we have $\|{\bf Z}_\alpha\hat{\bm\beta}_\alpha\|^2={\bm y}^\top{\bf Z}_\alpha({\bf Z}_\alpha^\top{\bf Z}_\alpha)^{-1}{\bf Z}_\alpha^\top{\bm y}\ge n^{-3/2+c_4}\|{\bf Z}_\alpha^\top{\bm y}\|^2=n^{1/2+c_4}\sum_{(j,k)\in\alpha}{\rm cor}^2({\bm z}_{jk},{\bm y})$, implying that the average of  
${\rm cor}^2({\bm z}_{jk},{\bm y})$ for $(j,k)\in\alpha$ is $o(n^{-c_4+c})$ if $|\alpha|=K=[\gamma n]$ for any $c>0$. By Condition 2, if $c_2+c_4>1/2$, then the average is lower than ${\rm cor}^2({\bm z}_{jk},{\bm y})$ for any $(j,k)\in\alpha^*$, implying that the set of the largest $[\gamma{n}]$ absolute correlation coefficients contains all of the important main and interaction effects in probability. We obtain the conclusion. \qed

\noindent
{\bf Proof of Theorem~\ref{thm:polynomial increases of p}.} Using the same method in the proof of Theorem~\ref{thm:consistency of SIS}, we can show that $\max_{|\alpha|=K}K^{-1}\|{\bf Z}_\alpha\hat{\bm\beta}_\alpha\|^2=o(n^{c})$ for any $c>0$. We then follow the remaining part of the proof of Theorem~\ref{thm:consistency of SIS} and obtain the conclusion. \qed

\noindent
{\bf Proof of Theorem~\ref{thm:oracle properties}.}  Conclusion (a) can be directly implied by the combination of Corollary 1 of~\cite{fantang2013} with Theorems~\ref{thm:consistency of SIS} or~\ref{thm:polynomial increases of p} for the case when $p$ exponentially or polynomially grows with $n$. Using Conclusion (a) and the theoretical properties of the SCAD and MCP discussed by~\cite{fanli2001} and~\cite{zhang2010}, respectively, we draw Conclusion (b). \qed

\noindent
{\bf Proof of Theorem~\ref{thm:consistency o the LASSO}}. The conclusion can be implied by the combination of Theorems~\ref{thm:consistency of SIS} or~\ref{thm:polynomial increases of p} with the well-known asymptotic results for the LASSO, where a review of those can be found in \cite{wu2020}.

\end{document}